\documentclass[20pt]{article}
\usepackage{amsmath}
\usepackage{amssymb}
\usepackage{latexsym}
\usepackage{amsmath}
\usepackage{lineno}
\usepackage{graphicx}
\usepackage{float}
\usepackage{subfigure}
\usepackage{geometry}
\usepackage{graphics}
\usepackage{subfigure}
\usepackage{graphicx}
\usepackage{multirow}
\usepackage{booktabs}
\usepackage{hyperref}
\usepackage{cite}
\makeatletter

\newcommand{\be}{\begin{equation}}
\newcommand{\ee}{\end{equation}}

\begin{document}
\begin{center}
\large {\bf Constraining the generalized uncertainty principle with the gravitational wave event GW150914}
\end{center}

\begin{center}
Zhong-Wen Feng $^1$
 $\footnote{E-mail:  \texttt{zwfengphy@163.com}}$
Shu-Zheng Yang $^2$
 $\footnote{E-mail:  \texttt{szyangcwnu@126.com}}$
Hui-Ling Li $ ^{1, 3}$
 $\footnote{E-mail: \texttt{LHL51759@126.com}}$
Xiao-Tao Zu ${^1}$
 $\footnote{E-mail:  \texttt{xtzu@uestc.edu.cn}}$
\end{center}

\begin{center}
\textit{1. School of Physical Electronics, University of Electronic Science and Technology of China, Chengdu, 610054, China\\
2. Department of Astronomy, China West Normal University, Nanchong, 637009, China\\
3. College of Physics Science and Technology, Shenyang Normal University, Shenyang, 110034, China}
\end{center}
\noindent
{\bf Abstract:} In this letter, we show that the dimensionless parameter in the generalized uncertainty principle (GUP) can be constrained by the gravitational wave event GW150914, which was discovered by the LIGO Scientific and Virgo Collaborations. Firstly, according to the Heisenberg uncertainty principle (HUP) and the data of gravitational wave event GW150914, we derive the standard energy-momentum dispersion relation and calculate the difference between the propagation speed of gravitons and the speed of light, i.e., $\Delta \upsilon$. Next, using two proposals regarding the GUP, we also generalize our study to the quantum gravity case and obtain the modified speed of gravitons. Finally, based on the modified speed of gravitons and $\Delta \upsilon$, the improved upper bounds on the GUP parameters are obtained. The results show that the upper limits of the GUP parameters $\beta_0$ and $\alpha_0$ are $2.3 \times 10^ {60}$ and $1.8 \times 10^{20}$.

\section{Introduction}
\label{Int}
Several versions of quantum gravity models predict the existence of a fundamental scale of length, which can be identified with the Planck scale. This view is also supported by Gedanken experiments \cite{ch0}. In this scenario, the Heisenberg uncertainty principle (HUP) can be changed into the so-called generalized uncertainty principle (GUP), which has various implications for a wide range of physical systems~\cite{ch0a,ch0b,ch0c,ch1,ch2,ch3,ch4,ch5,ch6,ch7,ch8,ch9,ch10,ch11,ch12,ch13,ch14,ch14+,ch15+,ch16a,ch16b,ch16c,ch16+}. For example, the effects of the GUP on the evolution of black holes were calculated in~\cite{ch1,ch2,ch3,ch4,ch5,ch6,ch7,ch8,ch9,ch10}. The GUP-corrected quantum Hall effect was investigated in~\cite{ch11}. The impact of the GUP on neutrino oscillations was studied in~\cite{ch12}. In~\cite{ch13,ch14}, using the GUP, the authors discussed the thermodynamics of the Friedmann-Robertson-Walker universe and the inflation preheating in cosmology. The effect of the GUP has also been used to calculate the entropic force \cite{ch14+,ch15+}. According to the GUP, the critical temperature and Helmholtz free energy of Bose-Einstein condensation in the relativistic ideal Bose gas were computed in~\cite{ch16a}. In~\cite{ch16b,ch16c}, Faizal and Majumder incorporated the GUP into Lifshitz field theories and showed that the breaking of supersymmetry by a non-anticommutative deformation can be used to generate the GUP. Moreover, it is exciting that the effects of the GUP can be probed via quantum optics~\cite{ch16+}. Based on this previous research, it is clear that two types of GUP have been studied widely. The version of
the proposed by Kempf, Mangano, and Mann (GUP I) is
\begin{equation}
\label{eq1}
\Delta x\Delta p \ge \frac{\hbar }{2}\left[ {1 + \beta \left( {\Delta p} \right)^2 } \right],
\end{equation}
where  $\Delta x$  and $\Delta p$ are the uncertainties for position and momentum, respectively. $\beta  = {{\beta _0 \ell_p^2 } \mathord{\left/ {\vphantom {{\beta _0 \ell_p^2 } {\hbar ^2 }}} \right. \kern-\nulldelimiterspace} {\hbar ^2 }} = {{\beta _0 } \mathord{\left/ {\vphantom {{\beta _0 } {M_p^2 c^2 }}} \right. \kern-\nulldelimiterspace} {M_p^2 c^2 }}$, $\beta _0$  is a positive dimensionless parameter, which is called the GUP parameter,  $\ell_p$ represents the Planck length,  $M_p$ is the  Planck mass, and the Planck energy is $M_p c^2 = 1.2 \times 10^{28} {\rm{eV}}$. Eq.~(\ref{eq1}) implies a nonzero minimal uncertainty  $\Delta x_{\min }  \approx \ell_p \sqrt {\beta _0 }$. It should be noted that the derivation of Eq.~(\ref{eq1}) relies on the modified fundamental commutation relation $\left[ {x_i ,p_j } \right] = i\hbar \delta_{ij}\left[ {1 + \beta p^2 } \right]$  with the position operator $x_i$  and the momentum operator $p_i$~\cite{ch15}. The other version proposed in~\cite{ch16,ch17} admits a minimal length and a maximal momentum (GUP II)
\begin{eqnarray}
\label{eq2}
\Delta x\Delta p & \ge & \frac{\hbar }{2}\left[ {1 - 2\alpha \left\langle p \right\rangle  + 4\alpha ^2 \left\langle {p^2 } \right\rangle } \right]
\nonumber \\
& \ge & \frac{\hbar }{2}\left[ {1 + \left( {\frac{\alpha }{{\sqrt {\left\langle {p^2 } \right\rangle } }} + 4\alpha ^2 } \right)\Delta p^2   } \right.
+ \left. {4\alpha ^2 \left\langle p \right\rangle ^2 + 4\alpha ^2 \left\langle p \right\rangle ^2 -   2\alpha \sqrt {\left\langle {p^2 } \right\rangle } } \right],
 \end{eqnarray}
where  $\alpha  = {{\alpha _0 } \mathord{\left/ {\vphantom {{\alpha _0 } {M_p c}}} \right. \kern-\nulldelimiterspace} {M_p c}} = {{\alpha _0 \ell_p } \mathord{\left/ {\vphantom {{\alpha _0 l_p } \hbar }} \right. \kern-\nulldelimiterspace} \hbar }$, and $\alpha_0$  is the GUP parameter. Inequality (\ref{eq2}) is equivalent to the modified fundamental commutation relation  $[x_i ,p_i ] = i\hbar [\delta _{ij}  - \alpha (p\delta _{ij}  + {{p_i p_j } \mathord{\left/ {\vphantom {{p_i p_j } p}} \right. \kern-\nulldelimiterspace} p}) + \alpha ^2 (p^2 \delta _{ij}  + 3p_i p_j )]$. Moreover, Eq.~(\ref{eq1}) implies a minimal length $\Delta x_{\min }  \approx \alpha _0 \ell_p$ and a maximum momentum  $\Delta p_{\max }  \approx {{\ell_p } \mathord{\left/ {\vphantom {{\ell_p } {\alpha _0 }}} \right. \kern-\nulldelimiterspace} {\alpha _0 }}$.

Theoretically, the GUP parameters $\beta_0$  and $\alpha_0$ are always assumed to be of the order of unity, in which case corrections are negligible unless the lengths or energies approach the Planck length $\ell_p$ or the Planck energy $E_p$. However, if the assumptions regarding the GUP parameters are not made a priori, the GUP parameters should be constrained by previous experiments~\cite{ch18,ch18+,ch18a+,ch19,ch19+,ch20+,ch20a+,ch21+,ch22+}. Recently, it was argued that the modified dispersion relations motivated by quantum gravity can affect the propagation of the observed gravitational wave signal \cite{ch23+}. Therefore, one can use the data of gravitational waves event GW150914 that reported by the Laser Interferometer Gravitational-Wave Observatory (LIGO) Scientific and Virgo Collaborations to constrain the possibility of Lorentz violation during gravitational waves propagation. In~\cite{ch20}, the authors showed that the upper bound on the difference between the speed of light and gravitational waves is  $ \left| {\Delta \upsilon } \right| \le 10^{ - 17}$. According to the gravitational wave event GW150914, Arzano and Calcagni showed that the upper bound on the characteristic quantum-gravity mass scale is  $M > 4 \times 10^4 {\rm{eV}}$, which is much weaker than that coming from photon propagation from gamma-ray bursts. They also found that the a phenomenological dispersion relation $\omega ^2  = k^2 \left( {1 + {{\alpha k^n } \mathord{\left/ {\vphantom {{\alpha k^n } {M^n }}} \right.  \kern- \nulldelimiterspace} {M^n }}} \right)$ is compatible with observations and it has a phenomenologically viable mass of $M > 10 {\rm{TeV}}$  only in the quite restrictive range of  $0<n<0.68$ \cite{ch21}. In order to explain why the speed of gravitational waves is smaller than that of light, Gwak, Kim, and Lee studied the speed of gravitons in the gravitational wave event GW150914 by using gravity's rainbow \cite{ch22}. They noted that the upper bound of the rainbow parameter $\eta$ is smaller than  $4.6 \times 10^{59}$ at a frequency of $250 {\rm{Hz}}$, which indicates that the effect of gravity's rainbow is very small in the gravitational waves event GW150914.

On the other hand, it is well known that the GUP is also motivated by the quantum gravity. Thus, inspired by previous studies, it is beneficial to investigate the possibility of using the gravitational wave event GW150914 to constrain the GUP parameters $\beta_0$ and $\alpha_0$. In this study, we first calculate the difference between the speed of gravitons and the speed of light via the HUP, i.e., $\Delta \upsilon$. Next, according to Eq.~(\ref{eq1}) and Eq.~(\ref{eq2}), we obtain the corresponding modified dispersion relations and the GUP-corrected speed of gravitons. Finally, we find the improved upper bounds on the GUP parameters by comparing $\Delta \upsilon$ with the GUP corrected speed of gravitons.

The remainder of this letter is organized as follows. In Section~\ref{sec2}, we derive the standard energy-momentum dispersion relation that corresponding to the HUP. Subsequently, following the definition of the group speed of the wave front, the difference between the speed of gravitons and  the speed of light is obtained. In Section~\ref{sec3} and Section~\ref{sec4}, we generalize our study to the quantum gravity case and compute the effects of GUP on the speed of gravitons. According to the calculated effect of the GUP on the speed of gravitons measurements, we obtain the improved upper bounds on the GUP parameters $\beta_0$ and $\alpha_0$. Finally, in Section~\ref{sec5} results are briefly discussed.

\section{The speed of gravitons in the Gravitational waves event GW150914}
\label{sec2}
First, we consider the conventional Heisenberg uncertainty principle $\Delta x\Delta p \ge {\hbar  \mathord{\left/ {\vphantom {\hbar  2}} \right. \kern-\nulldelimiterspace} 2}$. This inequality is equivalent to the Heisenberg algebra  $\left[ {x_i ,p_j } \right] = i\hbar \delta _{ij}$. The the position and momentum operators in the Heisenberg algebra can be defined as follows
\begin{equation}
\label{eq3}
\begin{array}{*{20}c}
   {x_i  = x_{0i}; \quad  p_i  = p_{0i}}, \\
\end{array}
\end{equation}
where $x_{0i}$ and $p_{0i}$ satisfying the canonical commutation relations $\left[ {x_{0i} ,p_{0j} } \right] = i\hbar \delta _{ij}$. In gravitational spacetime, the background metric ansatz that we study is
\begin{equation}
\label{eq4}
ds^2  = g_{ab} dx^a dx^b  = g_{00} c^2 dt^2  + g_{ij} dx^i dx^j .
\end{equation}
Eq.~(\ref{eq4}) leads to the square of the four-momentum  $p_a p^a  = g_{ab} p^a p^b  = g_{00} \left( {p^{0} } \right)^2  + g_{ij} p^{0i} p^{0j}$. Considering that $p_i$ in this background satisfies the relation  $p^2  = p_i p^i  = g_{ij} p^{0i} p^{0j}$, then the square of the four-momentum can be rewritten as
\begin{equation}
\label{eq5}
p_a p^a  = g_{00} \left( {p^0 } \right)^2  + p^2 .
\end{equation}
It should be mentioned that the right-hand side in RHS of Eq.~(\ref{eq5}) forms the original dispersion relation  $ g_{00} \left( {p^0 } \right)^2  + p^2  =  - m^2 c^2$. Therefore, the equation above takes the form
\begin{equation}
\label{eq6}
p_a p^a  =  - m^2 c^2 ,
\end{equation}
and the time component of the momentum is given by
\begin{equation}
\label{eq7}
\left( {p^0 } \right)^2  = {{\left( { - p^2  - m^2 c^2 } \right)} \mathord{\left/
 {\vphantom {{\left( { - p^2  - m^2 c^2 } \right)} {g_{00} }}} \right.
 \kern-\nulldelimiterspace} {g_{00} }}.
\end{equation}
It is well known that the energy of a particle can be defined in the following form
\begin{equation}
\label{eq8}
{\omega  \mathord{\left/
 {\vphantom {\omega  c}} \right.
 \kern-\nulldelimiterspace} c} =  - \xi _a p^a  =  - g_{ab} \xi ^a p^b,
\end{equation}
where  $\xi ^a  = \left( {1,0,0, \cdots } \right)$ is the Killing vector. Using Eq.~(\ref{eq8}), the energy in the metric~(\ref{eq4}) can be defined as \cite{ch23}
\begin{equation}
\label{eq9}
\omega  =  - g_{00} cp^0.
\end{equation}
Now, by substituting Eq.~(\ref{eq9}) into Eq.~(\ref{eq7}), one can express the energy of a particle in terms of the three spatial momentum momentum and the mass
\begin{equation}
\label{eq10}
\omega ^2  = \left( { - g_{00} cp^0 } \right)^2  =  - g_{00} \left( {p^2 c^2  + m^2 c^4 } \right).
\end{equation}
It is well known that the LIGO is in a weak gravitational spacetime, here we only work in the Minkowski spacetime, i.e.,  $g_{00}  =  - 1$. Thus, Eq.~(\ref{eq10}) reduces to the standard energy-momentum dispersion relation  $\omega ^2  = p^2 c^2  + m^2 c^4$. In addition, by assuming that the gravitational waves propagate as free waves, the speed of gravitons can be calculated by using the group speed of the wave front, i.e., $\upsilon : =  {{\partial \omega } \mathord{\left/  {\vphantom {{\partial \omega } {\partial p}}} \right. \kern-\nulldelimiterspace} {\partial p}}$, where  $\omega$ and $p$ represent the energy and momentum, respectively~\cite{ch23+,ch24+}. For the standard energy-momentum dispersion relation, the speed of gravitons is given by
\begin{eqnarray}
\label{eq11}
\upsilon _g  = \frac{{\partial \omega }}{{\partial p}} = \frac{{c^2 p}}{{\sqrt {c^4 m^2  + c^2 p^2 } }}
=  c\sqrt {1 - \frac{{m^2 c^4 }}{{\omega_g ^2 }}} \approx c\left( {1 - \frac{{m_g^2 c^4 }}{{2\omega _g^2 }}} \right),
\end{eqnarray}
where  $\omega _g$ and $m _g$ are the energy and rest mass of gravitons, respectively. Considering that $h = 4.136 \times 10^{ -15}{\rm{eV}} \cdot {\rm{s}}$ and $c = 3 \times 10^{8} {{\rm{m}} \mathord{\left/ {\vphantom {{\rm{m}} {\rm{s}}}} \right. \kern-\nulldelimiterspace} {\rm{s}}}$, then the difference between the speed of gravitons $\upsilon _g$ and the speed of light $c$ can be expressed as
\begin{equation}
\label{eq12+}
{\Delta \upsilon } = c - \upsilon _g = {{m_g^2 c^5 } \mathord{\left/ {\vphantom {{m_g^2 c^5 } {2\omega _g^2 }}} \right. \kern-\nulldelimiterspace} {2\omega _g^2 }}.
\end{equation}
In~\cite{ch24,ch25}, the LIGO Scientific and Virgo Collaborations pointed out that the signal of gravitational wave event GW150914 is peaked at $\nu= 150{\rm{Hz}}$. Thus, the maximum energy of gravitons is $\omega _g = h \nu \approx 6.024 \times 10^{ - 13} {\rm{eV}}$. Moreover, they also found an upper bound for the mass of gravitons is $m_g  \le 1.2 \times 10^{ - 22} {{{\rm{eV}}} \mathord{\left/ {\vphantom {{{\rm{eV}}} {c^2 }}} \right.  \kern-\nulldelimiterspace} {c^2 }}$. Therefore, Eq.~(\ref{eq12+}) turns out to be
\begin{equation}
\label{eq12}
 {\Delta \upsilon } < 5.6 \times 10^{ - 12} {\rm{m/s}}.
\end{equation}
It is clear that the difference between the speed of gravitons and the speed of light is very small. However, by using Eq.~(\ref{eq12}), one can investigate the appropriate range of the parameter in GUP motivated by quantum gravity. In Section~\ref{sec3} and Section~\ref{sec4}, by considering the GUP corrected speed of gravitons that obtained from the gravitational wave event GW150914 and Eq.~(\ref{eq12}), the upper limit of GUP parameters will be calculated.

\section{Bounds on the GUP parameters $\beta_0$}
\label{sec3}
In this section, we set the upper bound of the GUP parameter $\beta_0$ by considering the speed obtained from the gravitational wave event GW150914 and Eq.~(\ref{eq12}). In Eq.~(\ref{eq1}), the operators for the position $x_i$  and momentum $p_i$  can be defined as
\begin{eqnarray}
\label{eq13}
x_i  = x_{0i}; \quad p_i  = p_{0i} \left( {1 + \beta p^2 } \right),
\end{eqnarray}
where $x_{0i}$, $p_{0i}$ satisfying the canonical commutation relations $\left[ {x_{0i} ,p_{0j} } \right] = i\hbar \delta _{ij}$, and $p^2  = g_{ij} p^{0i} p^{0j}$, leading to $p = \sqrt {g_{ij} p^{0i} p^{0j}}$. Moreover, the $p_{0i}$ represents the the momentum at low energies, which satisfies the standard representation in position space, i.e., $p_{0i}  = {{ - i\hbar d} \mathord{\left/ {\vphantom {{ - i\hbar d} {dx_i }}} \right. \kern-\nulldelimiterspace} {dx_i }}$, and the $p_{i}$ as that at lower energies. In the Minkowski spacetime, the modified square of the four-momentum is given by
\begin{eqnarray}
\label{eq14}
p_a p^a  =  g_{ab} p^a p^b
=  - g_{00} \left( {p^{0} } \right)^2  + g_{ij} p^{0i} p^{0j} \left( {1 + \beta p^2 } \right)^2.
\end{eqnarray}
Then, by retaining the terms up to $\mathcal{O}\left(\beta^2\right)$, the equation above can be rewritten as
\begin{equation}
\label{eq15}
p_a p^a  =  - \left( {p^{0} } \right)^2  + p^2  + 2\beta p^2 p^2 ,
\end{equation}
 It should be noted that the first two terms of Eq.~(\ref{eq15}) form the usual dispersion relation, i.e., $- \left( {p^0 } \right)^2  + p^2 = - m^2 c^2$. Therefore, Eq.~(\ref{eq15}) can be rewritten as  $p_a p^a  = - m^2 c^2  + 2\beta p^2 p^2 $, and the time component of the momentum becomes
\begin{equation}
\label{eq16}
\left( {p^0 } \right)^2  = m^2 c^2 + p^2 \left( {1 - 2\beta p^2 } \right).
\end{equation}
According to Eq.~(\ref{eq8}) and Eq.~(\ref{eq9}), the energy of a particle can be expressed in terms of the three spatial momentum momentum and mass as follows
\begin{equation}
\label{eq17}
\omega ^2  = m^2 c^4 + p^2 c^2 \left( {1 - 2\beta p^2 } \right).
\end{equation}
Note that as  $\beta  \to 0$, Eq.~(\ref{eq17}) goes to the standard energy-momentum dispersion relation. Next, using Eq.~(\ref{eq11}), the group speed of massless gravitons with a modified dispersion relation~(\ref{eq17}) is approximately equal to
\begin{equation}
\label{eq17+}
\upsilon _{massless}  = \frac{{\partial \omega }}{{\partial p}} =\frac{{c\left( {1 - 4p^2 \beta } \right)}}{{\sqrt {1 - 2p^2 \beta } }} \approx c\left( {1 - 3\beta p^2 } \right).
\end{equation}
In the infrared, one can make a correction to the massless dispersion relation  $\omega_g ^2  = p_g^2 c^2$ \cite{ch21}. Thus, the Eq.~(\ref{eq17+}) can be rewritten as
\begin{equation}
\label{eq18}
\upsilon _{massless}  \approx c\left( {1 - 3\beta {{\omega_g ^2 } \mathord{\left/ {\vphantom {{\omega ^2 } {c^2 }}} \right. \kern-\nulldelimiterspace} {c^2 }}} \right),
\end{equation}
where $\omega_g$ is the energy of gravitons. The difference between the modified speed of gravitons and the speed of light is
\begin{equation}
\label{eq19}
\Delta \upsilon '= c-\upsilon _{massless} = {{3\beta \omega_g ^2 } \mathord{\left/ {\vphantom {{3\beta \omega_g ^2 } c}} \right. \kern-\nulldelimiterspace} c}
= {{3\beta _0 \omega_g ^2 } \mathord{\left/ {\vphantom {{3\beta _0 \omega_g ^2 } {M_p^2 c^3 }}} \right. \kern-\nulldelimiterspace} {M_p^2 c^3 }}.
\end{equation}
From the equation above, it is clear that the GUP slows down the speed of gravitons. In our previous work \cite{ch15+}, we assumed that $\beta_0\sim 1$. However, if this assumption is made, this will lead to a non-zero but virtually unmeasurable effect of the GUP. Conversely, without this assumption, one can set an improved upper bound on the GUP parameter by comparing $\Delta \upsilon '$  with Eq.~(\ref{eq12}), which is
\begin{equation}
\label{eq19+}
 {{3\beta _0 \omega_g ^2 } \mathord{\left/ {\vphantom {{3\beta _0 \omega_g ^2 } {M_p^2 c^3 }}} \right. \kern-\nulldelimiterspace} {M_p^2 c^3 }}< 5.6 \times 10^{ - 12} {\rm{m/s}}.
\end{equation}
Since the signal of gravitational wave event GW150914 is peaked at $ 150{\rm{Hz}}$, here we set $\omega _g  \approx 6.024 \times 10^{ - 13} {\rm{eV}}$. Therefore, the upper bound on the GUP parameter $\beta _0$ turns out to be
\begin{equation}
\label{eq20}
\beta _0  < 2.3 \times 10^ {60}.
\end{equation}
By comparing our upper bound on $\beta_0$ with the results in Table.~\ref{tab1}, it is easy to see that our result is weaker than those set by the tunneling current in a scanning tunneling microscope, the position measurement, the hydrogen lamb shift, the $^{87}$Rb cold-atom-recoil experiment and the Landau levels, whereas it is more stringent than those derived in a previous work~\cite{ch20a+}. However, it is believe that the accuracy of $\beta_0$ will get increased by several orders of magnitude with more accurate measurements in the further.

\begin{table}[htbp]
\caption {\label{tab1} Current experimental bounds on the GUP parameter $\beta_0$}
\centering
\begin{tabular}{c c c c c}
\toprule
Measurement/Experiment                 &  $\beta_0 $         &   Refs. &\\
\midrule
Electron tunneling                      & $10^{21}$          &  \cite{ch18,ch18+} &  \\
Gravitational bar detectors             & $10^{33}$          &  \cite{ch18a+}     &  \\
Electroweak measurement                    & $10^{34}$          &  \cite{ch18,ch18+} &  \\
Lamb shift                              & $10^{36}$          &  \cite{ch18,ch18+} &  \\
$^{87}$Rb cold-atom-recoil experiment   & $10^{39}$          &  \cite{ch21+}       &  \\
Landau levels                           & $10^{50}$          &  \cite{ch18,ch18+} &  \\
Perihelion precession
\\
(Solar system data)     & $10^{69}$          &  \cite{ch20a+}     &  \\
Perihelion precession
\\
(Pulsar PRS B 1913+16 data)& $10^{71}$       &  \cite{ch20a+}     &  \\
Modified mass-temperature relation      & $10^{78}$          &  \cite{ch20a+}     &  \\
Light deflection                        & $10^{78}$          &  \cite{ch20a+}     &  \\
\bottomrule
\end{tabular}
\end{table}

\section{Bounds on the GUP parameters $\alpha_0$}
\label{sec4}
In this section we apply the formalism given above to the GUP II case, so the effect of the GUP-induced term can be measurable. Now, let us define the operators for the position $x_i$  and momentum $p_i$ in GUP II as follows
\begin{eqnarray}
\label{eq21}
x_i  = x_{0i}; \quad p_i  =  p_{0i} \left( {1 - \alpha p  + 2\alpha ^2 p^2 } \right),
\end{eqnarray}
where $p^2  = g_{ij} p^{0i} p^{0j}$, thus $p = \sqrt {g_{ij} p^{0i} p^{0j}}$, $x_{0i}$ and $p_{0i}$ satisfying the canonical commutation relations $\left[ {x_{0i} ,p_{0j} } \right] = i\hbar \delta _{ij}$. The $p_{0i}$ is the the momentum at low energies (having the standard representation in position space $p_{0i}  = {{ - i\hbar d} \mathord{\left/ {\vphantom {{ - i\hbar d} {dx_i }}} \right. \kern-\nulldelimiterspace} {dx_i }}$), and the $p_{i}$ as that at lower energies. Similar to the GUP I case, the modified dispersion relation is derived as~\cite{ch26}
\begin{equation}
\label{eq22}
\omega ^2  = m^2 c^4 + p^2 c^2 \left( {1 - \alpha p} \right)^2 .
\end{equation}
Using Eq.~(\ref{eq8}) and the massless dispersion relation $\omega ^2  = p^2 c^2$, and by ignoring the higher order term of $\mathcal{O}\left( \alpha  \right)$, the speed of massless gravitons is
\begin{equation}
\label{eq22+}
\tilde \upsilon _{massless}  = c\left( {1 - {{2\alpha \omega_g } \mathord{\left/ {\vphantom {{2\alpha \omega_g } c}} \right. \kern-\nulldelimiterspace} c}} \right),
 \end{equation}
where $\omega_g$ is the graviton energy. Therefore, difference between the modified speed of gravitons and the speed of photon is given by
\begin{equation}
\label{eq23}
 {\Delta \upsilon }''  = c - \tilde \upsilon _{massless}  = {2\alpha \omega_g } < 5.6 \times 10^{ - 12} {\rm{m/s}}.
\end{equation}
From the inequality stated above, it is easy to see that the GUP corrected speed of gravitons is smaller than the speed of light. In order to obtained the upper bound on the GUP parameter, we need to compare the ${\Delta \upsilon }''$ with  Eq.~(\ref{eq12}), the result reads
\begin{equation}
\label{eq23+}
\alpha  < 4.5.
\end{equation}
In Eq.~(\ref{eq23+}), we still set $\omega _g  = 6.024 \times 10^{ - 13} {\rm{eV}}$ at a frequency of $\nu = 150 {\rm{Hz}}$. Next, according to Eq.~(\ref{eq2}), the upper bound on the GUP parameter $\alpha_0$ is given by
\begin{equation}
\label{eq24}
\alpha _0  < 1.8 \times 10^{20}.
\end{equation}
It should be stressed that the GUP parameter  $\alpha_0$ is attached to a term linear in  $p$, whereas the GUP parameter $\beta_0$  is attached to a term quadratic in $p$. Therefore, we should compare $\left( {\alpha _0 } \right)^2$ with  $\left( {\beta _0 } \right)$. It is obvious that the upper bound on  $\left( {\alpha _0 } \right)^2$ is 20 orders better than that on  $\beta _0$, which indicates that the gravitational wave measurements are more suitable for studying effects of the GUP II proposal.

 Table.~\ref{tab2} shows the bounds on $\alpha_0$ set by high-energy physics and other experiments.
\begin{table}[htbp]
\caption {\label{tab2} Current experimental bounds on GUP parameter $\alpha_0$}
\centering
\begin{tabular}{c c c c c}
\toprule
Measurement/Experiment                 & $\alpha_0 $        &   Refs. &\\
\midrule
Anomalous magnetic moment of the muon   & $10^{8} $          &  \cite{ch11} &  \\
Lamb shift                              & $10^{10}$          &  \cite{ch19} &  \\
Electron tunneling                      & $10^{11}$          &  \cite{ch19} &  \\
$^{87}$Rb cold-atom-recoil experiment   & $10^{14}$          &  \cite{ch21+}&  \\
Electroweak measurement                    & $10^{17}$          &  \cite{ch19} &  \\
Superconductivity                       & $10^{17}$          &  \cite{ch11} &  \\
Landau levels                           & $10^{23}$          &  \cite{ch19} &  \\
\bottomrule
\end{tabular}
\end{table}

Now, compared with the results in Table.~\ref{tab2}, one can find that the improved upper bound in Eq.~(\ref{eq24}) is weaker than that set by the anomalous magnetic moment of the muon, electron tunneling, hydrogen Lamb shift, $^{87}$Rb cold-atom-recoil experiment, position measurement and superconductivity experiment, whereas it is more stringent than that set Landau levels. In addition, it seems that our result is exceeding the stringent bound set by electroweak scale which is around $10^{17}$. We think the reason for this is that the spins of the black holes could not be measured accurately, which leads to the energy of gravitons ($6.024 \times 10^{ - 13} {\rm{eV}}$) is far weaker than the energy of electroweak measurements ($240{\rm{GeV}}$). With more accurate measurements in the further, the upper bound on $\alpha_0$ will reduce significantly with time \cite{ch31,ch32}. Therefore, our result is compatible with the bound set by electroweak scale. In particular, the LIGO Scientific and Virgo Collaborations recently reported the second gravitational wave event, GW151226, with the peak amplitude of the gravitational wave train $\nu=450{\rm{Hz}}$ \cite{ch30}. Using the same method, we find that the energy of gravitons is $\omega '_g  = h\nu = 1.8612 \times 10^{ - 12} {\rm{eV}}$ and the difference between the gravitons and speed of light becomes $\Delta \upsilon ' = 6.2 \times 10^{ - 13}$, which leading to the improved upper bound $\alpha _0'  < 6.7 \times 10^{18}$. Obviously, $\alpha _0'$ is more stringent than $\alpha _0$.

\section{Conclusion}
\label{sec5}
In previous study, it was generally assumed that the GUP parameters are of the order of unity. However, this assumption would lead to the effect of the GUP too small to be measured. By contrast, if this assumption is not made, one can constrain the GUP parameters based on previous experiments. In this letter, we considered two proposals regarding the GUP and used the data from the gravitational wave event GW150914 to calculate the modified speed of gravitons. Finally, we obtained improved upper bounds on the GUP parameters $\beta_0$ and $\alpha_0$. The results showed that the modified speed of gravitons is related to the effect of the GUP. When the GUP parameters approach zero, the modifications reduce to the original cases. Furthermore, we obtained $10^{60}$-level upper bound on $\beta_0$ and $10^{20}$-level upper bound on $\alpha_0$. Comparing the two kinds of GUP parameters, it is easy to see that the gravitational wave measurements are more suitable for studying effects of the GUP II proposal. Furthermore, it was shown that the upper bounds on GUP parameters $\beta_0$ and $\alpha_0$ are weaker than that set by electroweak measurement but not incompatible with it. With more accurate measurements in the further, the upper bounds on $\beta_0$ and $\alpha_0$ will reduce significantly with time. Finally, it is interesting to note that many studies have predicted that pulsars (spinning neutron stars) are promising candidates for producing the gravitational wave signals~\cite{ch27,ch28,ch29}. If these theories are proved by the gravitational wave detectors, one could obtain many new results in the future.

\vspace*{3.0ex}
{\bf Acknowledgements}
\vspace*{1.0ex}

The authors thank Klaus Doerbecker, Fabio Scardigli, Roberto Casadio,  Karissa Liu, and the anonymous referees for helpful suggestions and enlightening comments, which helped to improve the quality of this paper. This work was supported by the Natural Science Foundation of China (Grant No. 11573022).

\end{document}